\documentclass[aps,prl,showpacs,twocolumn]{revtex4}
\usepackage[usenames,dvipsnames]{color}
\usepackage{epsfig}
\newcommand{\newton}{G_\mathrm{N}}
\newcommand{\Bohr}{\rho_\mathrm{B}}
\newcommand{\muBohr}{\rho_{\mathrm{B},\mu p}}
\newcommand{\fm}{~\mathrm{fm}}
\newcommand{\mum}{~\mu\mathrm{m}}
\newcommand{\nm}{~\mathrm{nm}}
\newcommand{\mev}{~\mathrm{meV}}
\newcommand{\tev}{~\mathrm{TeV}}
\newcommand{\lew}{l_\mathrm{EW}}
\newcommand{\MD}{M_\mathrm{D}}

\newcommand{\khz}{~\mathrm{kHz}}

\newcommand{\jcap}{\rm {JCAP}}

\begin{document}

\title{Can Large Extra Dimensions Solve the Proton Radius Puzzle?}
\author{Zhigang Li$^{1,2}$\footnote{Email: zgli@bao.ac.cn}, 
Xuelei Chen$^{1,3}$\footnote{Email: xuelei@cosmology.bao.ac.cn}
}

\affiliation{
$^1$ Key Laboratory of Optical astronomy, National Astronomical Observatories,
     Chinese Academy of Science, Beijing 100012, China \\
$^2$ Graduate University of Chinese Academy of Sciences, Beijing 100049, China \\
$^3$ Center of High Energy Physics, Peking University, Beijing 100871, China
}
 
\date{\today}

\begin{abstract}
The proton charge radius extracted from the recent muonic hydrogen 
spectroscopy \cite{antognini2013proton,2010Natur.466..213P} differs 
from the CODATA 2010 recommended value \cite{2012RvMP...84.1527M} 
by more than $4\%$ or $4.4 \sigma$. This discrepancy, dubbed as the 
``Proton Radius Puzzle'', is a big challenge to the Standard Model 
of particle physics, and has triggered a number of works on the 
quantum electrodynamic calculations recently. The proton radius 
puzzle may indicate the presence of an extra correction which 
enlarges the 2S-2P energy gap in muonic hydrogen. Here we explore 
the possibility of large extra dimensions which could modify the 
Newtonian gravity at small scales and lower the 2S state energy
while leaving the 2P state nearly unchanged. 
We find that such effect could be produced by four or more large 
extra dimensions which are allowed by the current constraints from 
low energy physics.

\end{abstract}

\maketitle

\section{Introduction} \label{sec:intro}

Recently the charge radius of the proton is precisely measured from 
the Lamb shift of muonic hydrogen ($\mu p$) 
\cite{antognini2013proton,2010Natur.466..213P}, which yields 
$r_p=0.84087(39)$ fm. However, this value differs by $4.4 \sigma$ 
from the value obtained via electronic hydrogen (H) and deuterium 
(D) spectroscopic data, which is $0.8758(77)$ fm. It deviates even 
more (by $7 \sigma$) from the CODADA-2010 recommended value of 
$0.8775(51)$ fm, which was obtained from a combination of H and D 
spectroscopic data and the electron-proton scattering data 
\cite{2012RvMP...84.1527M}. 
Given the great precision in such measurements, this discrepancy, 
dubbed as the ``Proton Radius Puzzle'', is a severe problem and 
has received wide attention among physicists. The discrepancy has 
several possible explanations \cite{2013arXiv1301.0905P}. 

Of course, there is always the possibility that there is some hidden 
systematic error and the experiments are not as accurate as claimed. 
However, many independent electron-proton scattering experiments 
\cite{1962PhRv..126.1183L,1963RvMP...35..335H,1974PhRvC..10.2111M,
1974PhRvC...9.2125M,2003PhLB..576...62S,1980NuPhA.333..381S,
2005PhRvC..72e7601B,
2011PhLB..705...59Z,2012EPJA...48..151L,2010PhRvL.105x2001B,
2010PhRvD..82k3005H,
2012PrPNP..67..473S} 
are in good agreement with each other up to an arbitrary form factor 
extrapolation to low momentum transfers. The various measurements of 
transition frequencies in hydrogen 
\cite{2012RvMP...84.1527M,1997PhRvL..78..440D,1999PhRvL..82.4960S,
2010PhRvL.104w3001P,2011PhRvL.107t3001P} also agree with each other.
On the other hand, the muonic hydrogen experiments are even more
convincing than these
\cite{antognini2013proton,2010Natur.466..213P,2011AnPhy.326..516J,
2012PhRvL.109j3401K,
2006PhRvL..97s3402P}. 

On the theory side, a large missing term in the various quantum 
electrodynamic (QED) other than the proton structure effect, is 
highly unlikely, for the QED calculations in the hydrogen 
\cite{2001PhR...342...63I,2012RvMP...84.1527M,2004PhRvA..69f4103J,
2005PhRvA..72f2102J,2003PhRvL..91k3005P,1997PhRvA..55.3351E,
2010EPJD...58...57Y} and the muonic hydrogen 
\cite{2013arXiv1301.0905P,2012AnPhy.327..733B,2005PhRvA..71c2508B,
1982RvMP...54...67B,2012arXiv1208.2637A,2011AnPhy.326..500J,
2012arXiv1210.5828I,2011PhRvA..84a2506C,2001PhR...342...63I} 
have been done and checked by many groups over years. There is little 
room for a missing term or mistake which could contribute as large as 
$0.3$ meV. 

Recently, the proton polarizability contribution to Lamb shift in 
muonic hydrogen 
\cite{1996PhRvA..53.2092P,1999PhRvA..60.3593P,2006PAN....69.1309M}, 
which enters in the two-photon exchange term and involves intersection 
between the QED and QCD has been reanalyzed 
\cite{2011PhRvA..84a2506C,2011PhRvL.107p0402H,2011PhRvA..84b0101M,
2012PhRvC..86f5201M,2013PhLB..718.1078M}.
The uncertainty in this calculation could be large and might be able 
to account for the proton radius puzzle if there is a lepton-proton 
interaction caused by high-momentum behavior of the virtual scattering 
amplitude, which is proportional to the lepton mass to the fourth 
power. However, this assumption still needs to be confirmed in the 
future experiments. 

The possibility that the proton radius puzzle could imply the 
existence of a different interaction between $ep$ and $\mu p$ has 
been extensively explored, since such an interaction goes beyond the 
Standard Model (SM) of particle physics. The window for such new physics 
is small because the lepton universality has been rigorously tested 
\cite{2011PhRvL.106o3001B,2012PhRvL.108h1802B}. Several models, which 
assumed new gauge particles which couple the muon and/or proton with 
fine-tuned particle coupling or mass, have been proposed to reproduce 
the experiment results 
\cite{2011PhRvL.106o3001B,2012PhRvL.108h1802B,2011PhRvD..83c5020B,
2011PhRvL.107a1803B,2012PhRvD..86c5013C,2011PhRvD..83j1702T}. 
The models with new physics are promising but still primitive, and 
further works are needed to construct a complete gauge theory which 
can pass through all experiments.

The proton radius puzzle is still a puzzle. It is therefore valuable 
to search for new sources to explain the discrepancy of the extracted 
proton radius from hydrogen and muonic hydrogen experiments. In this 
paper, we explore whether large extra dimensions (LEDs) can solve the 
proton radius puzzle in the framework of a well-known low energy 
effective theory proposed by Arkani-Hamed, Dimopoulos and Dvali 
\cite{1998PhLB..429..263A} (ADD model). LEDs in ADD model can provide 
a stronger attractive gravitational field in scales much smaller than 
the size of them, which can lower the energy level of 2S state while 
leave 2P state nearly unchanged, so it could contribute to the Lamb 
shift. More importantly, muon has a mass about 206 times larger than the 
electron mass, so the Bohr radius of muonic hydrogen is about 186 times 
smaller than that of the hydrogen. This results in a contribution to Lamb 
shift in muonic hydrogen many orders of magnitude larger than that in 
hydrogen. Given appropriate number and size, which determine the 
strength of the modified gravitational field, LEDs could be a solution 
to the proton radius puzzle.

The ADD model was initially proposed to solve the hierarchical problem in 
Standard Model, i.e. the huge difference between the Planck scale of 
$\sim 10^{19}$ GeV and the electroweak scale ($\sim 1$ TeV), by assuming 
the existence of $n$ extra spatial dimensions of size $R_n$ which are 
compactified on an $n$-dimensional torus. The fundamental Planck scale 
in $(4+n)$-dimentional spacetime ($M_\mathrm{D}$) is then related to 
the $4$-dimensional effective Planck scale ($M_\mathrm{P}$) by Gauss's 
law, $M_\mathrm{P}^2=M_\mathrm{D}^{n+2}R_n^n$. The fundamental Planck 
scale can be tuned to $\sim 1$ TeV if the volume $\propto R_n^n$ 
is large enough, so the hierarchical problem is solved. One consequence 
of the ADD model is that at scales much smaller than the size of the 
extra dimensions $R_n$, the gravity force will be strengthened,  
$V(r)=V_\mathrm{N}(r) \times (R_n/r)^n$, 
where $V_\mathrm{N}(r)=G_\mathrm{N}m_1m_2/r$ is the Newtonian gravity 
between two particles with mass $m_1$ and $m_2$, and $G_\mathrm{N}$ 
is the Newton gravitational constant. 

We calculate the contribution of this modified gravity to the energy 
levels in hydrogen and muonic hydrogen in Sec. II, and give the size 
of extra-dimensions which can solve the proton radius puzzle. 
In Sec. III, we discuss the various constraints on the number and 
size of the LEDs.

\section{Lamb shift and proton radius puzzle}  \label{sec:ls}

If $n$ compactified LEDs are introduced, each with size of $R_n$, 
the gravity force between particles is modified on small scales as,
\begin{equation}
V(r) = \left\{
\begin{array}{ll}
\frac{\newton m_1m_2}{r} \times (\frac{R_n}{r})^n, &
         ~~~ \mathrm{for} ~ r \ll R_n  \\
      \frac{\newton m_1m_2}{r}, & ~~~ \mathrm{for} ~ r \gg R_n
\end{array}\right.
\label{eq:add}
\end{equation}
This may be treated as an perturbation on the standard Columb potential.
The correction on the energy levels is given by 
\begin{equation}
\Delta E_{nl} = \langle nl|V|nl \rangle,
\end{equation} 
where $n$ is the principal quantum number and $l=0,1,\ldots,n-1$ is the 
angular momentum quantum number. In the calculation of the perturbations 
on the hydrogen and muonic hydrogen energy levels, either the 
non-relativistic Sch\"odinger wave function or the relativistic Dirac 
wave function can be used, and for our purpose the two approaches give 
negligible small difference. Below we use the simpler Sch\"odinger wave 
functions to illustrate the calculation, but give the final results 
using the Dirac wave functions. 

Specifically, the LEDs contribute to the 2S state is reduced to
\begin{eqnarray}
\Delta E_\mathrm{2S} 
  &\approx& \frac{\newton m_l m_p}{8\Bohr} (\frac{R_n}{\Bohr})^n \ 
            \int_{t_0}^{t_n} \frac{(2-t)^2e^{-t}}{t^{n-1}} dt \nonumber \\
  &\approx& \frac{E_0}{2(n-2)} (\frac{R_n}{\Bohr})^n (\frac{r_0}{\Bohr})^{2-n}
\label{eq:2S}
\end{eqnarray}
where $E_0=\newton m_lm_p/\Bohr$, $\Bohr\equiv \hbar^2/m_r e^2$ is the 
Bohr radius with $m_r$ the reduced mass, $m_l$ the lepton (electron or 
muon) mass, and $m_p$ the proton mass. The lower limit of the integral 
$t_0=r_0/\Bohr$ is the small scale cutoff in unit of Bohr radius, and 
the upper limit is given by $t_n=R_n/\Bohr$. Eq.~(\ref{eq:2S}) is valid 
only if $n\ge3$. We may neglect the contribution of the Newtonian gravity 
at large scales. The Bohr radius for hydrogen is $\Bohr=0.529\times10^5\fm$, 
and for the muonic hydrogen it is $\muBohr=285\fm$, so if the electroweak 
scale ($l_\mathrm{EW}\approx2\times 10^{-3} \fm$) or even the proton 
radius ($r_p\approx 0.8 \fm$) is taken as the cutoff, $t_0 \sim 0$, and 
the second line of Eq.~(\ref{eq:2S}) results.  We see that the correction 
$\Delta E_\mathrm{2S} \propto m_l/\Bohr^3$,  so for the muonic hydrogen 
it is more than 9 orders of magnitude larger than for hydrogen.

Similarly, the correction of LEDs to the energy level of 2P state can 
be calculated as,
\begin{eqnarray}
\Delta E_\mathrm{2P} \approx E_0 \cdot (\frac{R_n}{\Bohr})^n \cdot \
      \frac{1}{24} \int_{t_0}^{t_n} \frac{e^{-t}}{t^{n-3}} dt,
\label{eq:2P}
\end{eqnarray}
which will be several orders of magnitude smaller than that of 
2S state and can be neglected in the following discussion.

Summing up all the contributions, the Lamb shift in the muonic 
hydrogen can be written as 
\cite{2012arXiv1208.2637A},
\begin{eqnarray}
\Delta E_\mathrm{LS}(\mev) = 206.0336(15)-5.2275(10)r^2_p+\Delta E_\mathrm{TPE},
\label{eq:lsth}
\end{eqnarray}
where $r_p$ is in units  of $\fm$.
The second term (mainly from the one-photon exchange and its radiative 
corrections) and the third term (two-photon exchange) are dependent on 
proton structure. The rms charge radius of proton can be extracted by 
comparing Eq.~(\ref{eq:lsth}) with the value from experiments. The 
difference between the proton radius extracted from the hydrogen and 
muonic hydrogen experiments results in a correction to the Lamb shift 
in muonic hydrogen,
\begin{eqnarray}
\delta E_\mathrm{LS,PRP} = 0.329(50)\mev
\label{eq:lsprp}
\end{eqnarray}
The error comes mainly from the hydrogen experiments.

Using Eqs.(\ref{eq:2S}-\ref{eq:lsprp}), we find that in order to 
explain the proton radius puzzle, the extra dimensions should have
\begin{eqnarray}
R_n = D_n \times (\frac{r_0}{\lew})^{1-2/n}
\label{eq:size}
\end{eqnarray}
where $D_n$ is the size of the $n$ large extra dimensions (here we 
take $r_0$ as the electroweak scale $\lew$). For $n=3,4,5,6$, 
we have 
$D_3=61\mum$, $D_4=31\nm$, $D_5=0.31\nm$, $D_6=0.014\nm$ respectively.

The modified gravity of LEDs as given above contributes $0.38\khz$ 
to the Lamb shift in hydrogen. This is compatible with the current 
measurements \cite{2012RvMP...84.1527M}. It also contributes 
$2.6\khz$ to the 1S-2S transition frequency, which is the best 
measured one in hydrogen spectroscopy, and this is 77 times larger 
than the CODATA 2010 suggested value of the experimental uncertainty 
\cite{2012RvMP...84.1527M}. However the theory of hydrogen energy 
level for 1S state has an uncertainty of the order of several kHz, 
mainly due to two-loop and three-loop calculations (see 
\cite{2013arXiv1301.0905P} and references therein). Taken into account
this theoretical uncertainty, the presence of LEDs with size shown in 
Eq.~(\ref{eq:size}) may still be allowed.

\section{Discussions}   \label{sec:sum}

In this work, we calculated the contribution to the Lamb shift of 
muonic hydrogen from the modified gravity force in models with large 
extra dimensions. To solve the proton radius puzzle, the LEDs should 
be as large as shown in Eq.~(\ref{eq:size}). 

There are many works on the constraint of the ADD model. 
The model with only one large extra dimension has been ruled out by 
the experiments over the solar system \cite{2004PhRvD..70d2004H}. 
The model with two or three LEDs are tightly constrained by the 
tests of the inverse-square law of gravity at short ranges using 
torsion pendulum or by measuring the Casimir force  
\cite{2007PhRvL..98b1101K,2011PhRvL.107q1101S,2012PhRvL.108h1101Y,
PhysRevD.78.022002,2012PhRvD..86f5025K}, $R_n\lesssim30\mum$. 
However, such experiments of testing non-Newtonian gravity do not 
have sufficient sensitivity when there are four or more LEDs 
\cite{2008ChPhB..17...70L,2007PhRvL..98m1104A}, since the ability 
to detect gravity at short ranges is limited by the closest distances 
the test mass can reach, which are typically of the order of several 
hundred nanometers, much larger than the required $R_n$ for $n\ge4$. 
Furthermore, the reliability of the Casimir force experiments, which 
can detect gravity at scales smaller than the torsion pendulum 
experiments, is still on debate \cite{2011LNP...834...97L} as there 
are complicated systematics such as the electric patch potential and 
thermal contributions to the Casimir forces.

Another problem with the ADD model is that it may have a Kaluza-Klein 
(KK) tower of gravitons propagating in the extra dimensions. The SM 
particles which are confined in the three dimensional space can be 
kicked off, leading to an abnormal missing of events with high energy 
transfer momentum $p\gtrsim\MD$. Absence of these phenomena in high 
energy physics experiments put the most stringent constraints on the 
fundamental Planck scale $M_\mathrm{D} \gtrsim 2 \tev$
\cite{2008PhDT.......280B,2011PhRvD..84a3003M,2012JCAP...02..012F, 
2013PhRvL.110a1802A,2012arXiv1211.1150A}, which would exclude the LED 
effect as a solution to the proton radius puzzle. However, such 
constraints depend on the nature of the KK gravitons which is not yet 
well understood. 
So the possible contribution of the ADD model to the Lamb shift should 
not be dismissed without further investigations.

Improvements on the measurements of the Lamb shift and the calculations 
of two-loop and three-loop corrections to the 1S and 2S energy levels in 
hydrogen are helpful to confirm or exclude the LEDs as the solution to 
the proton radius puzzle.

\acknowledgments

ZL and XC are supported by the National Science Foundation of China (NSFC) 
under grant No.11073024 and by the Ministry of Science and Technology under 
the 863 grant No.2012AA121701.


\begin{thebibliography}{64}
\expandafter\ifx\csname natexlab\endcsname\relax\def\natexlab#1{#1}\fi
\expandafter\ifx\csname bibnamefont\endcsname\relax
  \def\bibnamefont#1{#1}\fi
\expandafter\ifx\csname bibfnamefont\endcsname\relax
  \def\bibfnamefont#1{#1}\fi
\expandafter\ifx\csname citenamefont\endcsname\relax
  \def\citenamefont#1{#1}\fi
\expandafter\ifx\csname url\endcsname\relax
  \def\url#1{\texttt{#1}}\fi
\expandafter\ifx\csname urlprefix\endcsname\relax\def\urlprefix{URL }\fi
\providecommand{\bibinfo}[2]{#2}
\providecommand{\eprint}[2][]{\url{#2}}

\bibitem[{\citenamefont{Antognini et~al.}(2013)\citenamefont{Antognini, Nez,
  Schuhmann, Amaro, Biraben, Cardoso, Covita, Dax, Dhawan, Diepold
  et~al.}}]{antognini2013proton}
\bibinfo{author}{\bibfnamefont{A.}~\bibnamefont{Antognini}},
  \bibinfo{author}{\bibfnamefont{F.}~\bibnamefont{Nez}},
  \bibinfo{author}{\bibfnamefont{K.}~\bibnamefont{Schuhmann}},
  \bibinfo{author}{\bibfnamefont{F.~D.} \bibnamefont{Amaro}},
  \bibinfo{author}{\bibfnamefont{F.}~\bibnamefont{Biraben}},
  \bibinfo{author}{\bibfnamefont{J.~M.} \bibnamefont{Cardoso}},
  \bibinfo{author}{\bibfnamefont{D.~S.} \bibnamefont{Covita}},
  \bibinfo{author}{\bibfnamefont{A.}~\bibnamefont{Dax}},
  \bibinfo{author}{\bibfnamefont{S.}~\bibnamefont{Dhawan}},
  \bibinfo{author}{\bibfnamefont{M.}~\bibnamefont{Diepold}},
  \bibnamefont{et~al.}, \bibinfo{journal}{Science}
  \textbf{\bibinfo{volume}{339}}, \bibinfo{pages}{417} (\bibinfo{year}{2013}).

\bibitem[{\citenamefont{{Pohl} et~al.}(2010)\citenamefont{{Pohl}, {Antognini},
  {Nez}, {Amaro}, {Biraben}, {Cardoso}, {Covita}, {Dax}, {Dhawan}, {Fernandes}
  et~al.}}]{2010Natur.466..213P}
\bibinfo{author}{\bibfnamefont{R.}~\bibnamefont{{Pohl}}},
  \bibinfo{author}{\bibfnamefont{A.}~\bibnamefont{{Antognini}}},
  \bibinfo{author}{\bibfnamefont{F.}~\bibnamefont{{Nez}}},
  \bibinfo{author}{\bibfnamefont{F.~D.} \bibnamefont{{Amaro}}},
  \bibinfo{author}{\bibfnamefont{F.}~\bibnamefont{{Biraben}}},
  \bibinfo{author}{\bibfnamefont{J.~M.~R.} \bibnamefont{{Cardoso}}},
  \bibinfo{author}{\bibfnamefont{D.~S.} \bibnamefont{{Covita}}},
  \bibinfo{author}{\bibfnamefont{A.}~\bibnamefont{{Dax}}},
  \bibinfo{author}{\bibfnamefont{S.}~\bibnamefont{{Dhawan}}},
  \bibinfo{author}{\bibfnamefont{L.~M.~P.} \bibnamefont{{Fernandes}}},
  \bibnamefont{et~al.}, \bibinfo{journal}{\nat} \textbf{\bibinfo{volume}{466}},
  \bibinfo{pages}{213} (\bibinfo{year}{2010}).

\bibitem[{\citenamefont{{Mohr} et~al.}(2012)\citenamefont{{Mohr}, {Taylor}, and
  {Newell}}}]{2012RvMP...84.1527M}
\bibinfo{author}{\bibfnamefont{P.~J.} \bibnamefont{{Mohr}}},
  \bibinfo{author}{\bibfnamefont{B.~N.} \bibnamefont{{Taylor}}},
  \bibnamefont{and} \bibinfo{author}{\bibfnamefont{D.~B.}
  \bibnamefont{{Newell}}}, \bibinfo{journal}{Reviews of Modern Physics}
  \textbf{\bibinfo{volume}{84}}, \bibinfo{pages}{1527} (\bibinfo{year}{2012}),
  \eprint{1203.5425}.

\bibitem[{\citenamefont{{Pohl} et~al.}(2013)\citenamefont{{Pohl}, {Gilman},
  {Miller}, and {Pachucki}}}]{2013arXiv1301.0905P}
\bibinfo{author}{\bibfnamefont{R.}~\bibnamefont{{Pohl}}},
  \bibinfo{author}{\bibfnamefont{R.}~\bibnamefont{{Gilman}}},
  \bibinfo{author}{\bibfnamefont{G.~A.} \bibnamefont{{Miller}}},
  \bibnamefont{and}
  \bibinfo{author}{\bibfnamefont{K.}~\bibnamefont{{Pachucki}}},
  \bibinfo{journal}{ArXiv e-prints}  (\bibinfo{year}{2013}),
  \eprint{1301.0905}.

\bibitem[{\citenamefont{{Lehmann} et~al.}(1962)\citenamefont{{Lehmann},
  {Taylor}, and {Wilson}}}]{1962PhRv..126.1183L}
\bibinfo{author}{\bibfnamefont{P.}~\bibnamefont{{Lehmann}}},
  \bibinfo{author}{\bibfnamefont{R.}~\bibnamefont{{Taylor}}}, \bibnamefont{and}
  \bibinfo{author}{\bibfnamefont{R.}~\bibnamefont{{Wilson}}},
  \bibinfo{journal}{Physical Review} \textbf{\bibinfo{volume}{126}},
  \bibinfo{pages}{1183} (\bibinfo{year}{1962}).

\bibitem[{\citenamefont{{Hand} et~al.}(1963)\citenamefont{{Hand}, {Miller}, and
  {Wilson}}}]{1963RvMP...35..335H}
\bibinfo{author}{\bibfnamefont{L.~N.} \bibnamefont{{Hand}}},
  \bibinfo{author}{\bibfnamefont{D.~G.} \bibnamefont{{Miller}}},
  \bibnamefont{and} \bibinfo{author}{\bibfnamefont{R.}~\bibnamefont{{Wilson}}},
  \bibinfo{journal}{Reviews of Modern Physics} \textbf{\bibinfo{volume}{35}},
  \bibinfo{pages}{335} (\bibinfo{year}{1963}).

\bibitem[{\citenamefont{{Murphy}
  et~al.}(1974{\natexlab{a}})\citenamefont{{Murphy}, {Shin}, and
  {Skopik}}}]{1974PhRvC..10.2111M}
\bibinfo{author}{\bibfnamefont{J.~J.} \bibnamefont{{Murphy}}},
  \bibinfo{author}{\bibfnamefont{Y.~M.} \bibnamefont{{Shin}}},
  \bibnamefont{and} \bibinfo{author}{\bibfnamefont{D.~M.}
  \bibnamefont{{Skopik}}}, \bibinfo{journal}{\prc}
  \textbf{\bibinfo{volume}{10}}, \bibinfo{pages}{2111}
  (\bibinfo{year}{1974}{\natexlab{a}}).

\bibitem[{\citenamefont{{Murphy}
  et~al.}(1974{\natexlab{b}})\citenamefont{{Murphy}, {Shin}, and
  {Skopik}}}]{1974PhRvC...9.2125M}
\bibinfo{author}{\bibfnamefont{J.~J.} \bibnamefont{{Murphy}}},
  \bibinfo{author}{\bibfnamefont{Y.~M.} \bibnamefont{{Shin}}},
  \bibnamefont{and} \bibinfo{author}{\bibfnamefont{D.~M.}
  \bibnamefont{{Skopik}}}, \bibinfo{journal}{\prc}
  \textbf{\bibinfo{volume}{9}}, \bibinfo{pages}{2125}
  (\bibinfo{year}{1974}{\natexlab{b}}).

\bibitem[{\citenamefont{{Sick}}(2003)}]{2003PhLB..576...62S}
\bibinfo{author}{\bibfnamefont{I.}~\bibnamefont{{Sick}}},
  \bibinfo{journal}{Physics Letters B} \textbf{\bibinfo{volume}{576}},
  \bibinfo{pages}{62} (\bibinfo{year}{2003}), \eprint{arXiv:nucl-ex/0310008}.

\bibitem[{\citenamefont{{Simon} et~al.}(1980)\citenamefont{{Simon}, {Schmitt},
  {Borkowski}, and {Walther}}}]{1980NuPhA.333..381S}
\bibinfo{author}{\bibfnamefont{G.~G.} \bibnamefont{{Simon}}},
  \bibinfo{author}{\bibfnamefont{C.}~\bibnamefont{{Schmitt}}},
  \bibinfo{author}{\bibfnamefont{F.}~\bibnamefont{{Borkowski}}},
  \bibnamefont{and} \bibinfo{author}{\bibfnamefont{V.~H.}
  \bibnamefont{{Walther}}}, \bibinfo{journal}{Nuclear Physics A}
  \textbf{\bibinfo{volume}{333}}, \bibinfo{pages}{381} (\bibinfo{year}{1980}).

\bibitem[{\citenamefont{{Blunden} and {Sick}}(2005)}]{2005PhRvC..72e7601B}
\bibinfo{author}{\bibfnamefont{P.~G.} \bibnamefont{{Blunden}}}
  \bibnamefont{and} \bibinfo{author}{\bibfnamefont{I.}~\bibnamefont{{Sick}}},
  \bibinfo{journal}{\prc} \textbf{\bibinfo{volume}{72}}, \bibinfo{eid}{057601}
  (\bibinfo{year}{2005}), \eprint{arXiv:nucl-th/0508037}.

\bibitem[{\citenamefont{{Zhan} et~al.}(2011)\citenamefont{{Zhan}, {Allada},
  {Armstrong}, {Arrington}, {Bertozzi}, {Boeglin}, {Chen}, {Chirapatpimol},
  {Choi}, {Chudakov} et~al.}}]{2011PhLB..705...59Z}
\bibinfo{author}{\bibfnamefont{X.}~\bibnamefont{{Zhan}}},
  \bibinfo{author}{\bibfnamefont{K.}~\bibnamefont{{Allada}}},
  \bibinfo{author}{\bibfnamefont{D.~S.} \bibnamefont{{Armstrong}}},
  \bibinfo{author}{\bibfnamefont{J.}~\bibnamefont{{Arrington}}},
  \bibinfo{author}{\bibfnamefont{W.}~\bibnamefont{{Bertozzi}}},
  \bibinfo{author}{\bibfnamefont{W.}~\bibnamefont{{Boeglin}}},
  \bibinfo{author}{\bibfnamefont{J.-P.} \bibnamefont{{Chen}}},
  \bibinfo{author}{\bibfnamefont{K.}~\bibnamefont{{Chirapatpimol}}},
  \bibinfo{author}{\bibfnamefont{S.}~\bibnamefont{{Choi}}},
  \bibinfo{author}{\bibfnamefont{E.}~\bibnamefont{{Chudakov}}},
  \bibnamefont{et~al.}, \bibinfo{journal}{Physics Letters B}
  \textbf{\bibinfo{volume}{705}}, \bibinfo{pages}{59} (\bibinfo{year}{2011}),
  \eprint{1102.0318}.

\bibitem[{\citenamefont{{Lorenz} et~al.}(2012)\citenamefont{{Lorenz}, {Hammer},
  and {Mei{\ss}ner}}}]{2012EPJA...48..151L}
\bibinfo{author}{\bibfnamefont{I.~T.} \bibnamefont{{Lorenz}}},
  \bibinfo{author}{\bibfnamefont{H.-W.} \bibnamefont{{Hammer}}},
  \bibnamefont{and} \bibinfo{author}{\bibfnamefont{U.-G.}
  \bibnamefont{{Mei{\ss}ner}}}, \bibinfo{journal}{European Physical Journal A}
  \textbf{\bibinfo{volume}{48}}, \bibinfo{pages}{151} (\bibinfo{year}{2012}),
  \eprint{1205.6628}.

\bibitem[{\citenamefont{{Bernauer} et~al.}(2010)\citenamefont{{Bernauer},
  {Achenbach}, {Ayerbe Gayoso}, {B{\"o}hm}, {Bosnar}, {Debenjak}, {Distler},
  {Doria}, {Esser}, {Fonvieille} et~al.}}]{2010PhRvL.105x2001B}
\bibinfo{author}{\bibfnamefont{J.~C.} \bibnamefont{{Bernauer}}},
  \bibinfo{author}{\bibfnamefont{P.}~\bibnamefont{{Achenbach}}},
  \bibinfo{author}{\bibfnamefont{C.}~\bibnamefont{{Ayerbe Gayoso}}},
  \bibinfo{author}{\bibfnamefont{R.}~\bibnamefont{{B{\"o}hm}}},
  \bibinfo{author}{\bibfnamefont{D.}~\bibnamefont{{Bosnar}}},
  \bibinfo{author}{\bibfnamefont{L.}~\bibnamefont{{Debenjak}}},
  \bibinfo{author}{\bibfnamefont{M.~O.} \bibnamefont{{Distler}}},
  \bibinfo{author}{\bibfnamefont{L.}~\bibnamefont{{Doria}}},
  \bibinfo{author}{\bibfnamefont{A.}~\bibnamefont{{Esser}}},
  \bibinfo{author}{\bibfnamefont{H.}~\bibnamefont{{Fonvieille}}},
  \bibnamefont{et~al.}, \bibinfo{journal}{Physical Review Letters}
  \textbf{\bibinfo{volume}{105}}, \bibinfo{eid}{242001} (\bibinfo{year}{2010}),
  \eprint{1007.5076}.

\bibitem[{\citenamefont{{Hill} and {Paz}}(2010)}]{2010PhRvD..82k3005H}
\bibinfo{author}{\bibfnamefont{R.~J.} \bibnamefont{{Hill}}} \bibnamefont{and}
  \bibinfo{author}{\bibfnamefont{G.}~\bibnamefont{{Paz}}},
  \bibinfo{journal}{\prd} \textbf{\bibinfo{volume}{82}}, \bibinfo{eid}{113005}
  (\bibinfo{year}{2010}), \eprint{1008.4619}.

\bibitem[{\citenamefont{{Sick}}(2012)}]{2012PrPNP..67..473S}
\bibinfo{author}{\bibfnamefont{I.}~\bibnamefont{{Sick}}},
  \bibinfo{journal}{Progress in Particle and Nuclear Physics}
  \textbf{\bibinfo{volume}{67}}, \bibinfo{pages}{473} (\bibinfo{year}{2012}).

\bibitem[{\citenamefont{{de Beauvoir} et~al.}(1997)\citenamefont{{de Beauvoir},
  {Nez}, {Julien}, {Cagnac}, {Biraben}, {Touahri}, {Hilico}, {Acef}, {Clairon},
  and {Zondy}}}]{1997PhRvL..78..440D}
\bibinfo{author}{\bibfnamefont{B.}~\bibnamefont{{de Beauvoir}}},
  \bibinfo{author}{\bibfnamefont{F.}~\bibnamefont{{Nez}}},
  \bibinfo{author}{\bibfnamefont{L.}~\bibnamefont{{Julien}}},
  \bibinfo{author}{\bibfnamefont{B.}~\bibnamefont{{Cagnac}}},
  \bibinfo{author}{\bibfnamefont{F.}~\bibnamefont{{Biraben}}},
  \bibinfo{author}{\bibfnamefont{D.}~\bibnamefont{{Touahri}}},
  \bibinfo{author}{\bibfnamefont{L.}~\bibnamefont{{Hilico}}},
  \bibinfo{author}{\bibfnamefont{O.}~\bibnamefont{{Acef}}},
  \bibinfo{author}{\bibfnamefont{A.}~\bibnamefont{{Clairon}}},
  \bibnamefont{and} \bibinfo{author}{\bibfnamefont{J.~J.}
  \bibnamefont{{Zondy}}}, \bibinfo{journal}{Physical Review Letters}
  \textbf{\bibinfo{volume}{78}}, \bibinfo{pages}{440} (\bibinfo{year}{1997}).

\bibitem[{\citenamefont{{Schwob} et~al.}(1999)\citenamefont{{Schwob},
  {Jozefowski}, {de Beauvoir}, {Hilico}, {Nez}, {Julien}, {Biraben}, {Acef},
  {Zondy}, and {Clairon}}}]{1999PhRvL..82.4960S}
\bibinfo{author}{\bibfnamefont{C.}~\bibnamefont{{Schwob}}},
  \bibinfo{author}{\bibfnamefont{L.}~\bibnamefont{{Jozefowski}}},
  \bibinfo{author}{\bibfnamefont{B.}~\bibnamefont{{de Beauvoir}}},
  \bibinfo{author}{\bibfnamefont{L.}~\bibnamefont{{Hilico}}},
  \bibinfo{author}{\bibfnamefont{F.}~\bibnamefont{{Nez}}},
  \bibinfo{author}{\bibfnamefont{L.}~\bibnamefont{{Julien}}},
  \bibinfo{author}{\bibfnamefont{F.}~\bibnamefont{{Biraben}}},
  \bibinfo{author}{\bibfnamefont{O.}~\bibnamefont{{Acef}}},
  \bibinfo{author}{\bibfnamefont{J.-J.} \bibnamefont{{Zondy}}},
  \bibnamefont{and}
  \bibinfo{author}{\bibfnamefont{A.}~\bibnamefont{{Clairon}}},
  \bibinfo{journal}{Physical Review Letters} \textbf{\bibinfo{volume}{82}},
  \bibinfo{pages}{4960} (\bibinfo{year}{1999}).

\bibitem[{\citenamefont{{Parthey} et~al.}(2010)\citenamefont{{Parthey},
  {Matveev}, {Alnis}, {Pohl}, {Udem}, {Jentschura}, {Kolachevsky}, and
  {H{\"a}nsch}}}]{2010PhRvL.104w3001P}
\bibinfo{author}{\bibfnamefont{C.~G.} \bibnamefont{{Parthey}}},
  \bibinfo{author}{\bibfnamefont{A.}~\bibnamefont{{Matveev}}},
  \bibinfo{author}{\bibfnamefont{J.}~\bibnamefont{{Alnis}}},
  \bibinfo{author}{\bibfnamefont{R.}~\bibnamefont{{Pohl}}},
  \bibinfo{author}{\bibfnamefont{T.}~\bibnamefont{{Udem}}},
  \bibinfo{author}{\bibfnamefont{U.~D.} \bibnamefont{{Jentschura}}},
  \bibinfo{author}{\bibfnamefont{N.}~\bibnamefont{{Kolachevsky}}},
  \bibnamefont{and} \bibinfo{author}{\bibfnamefont{T.~W.}
  \bibnamefont{{H{\"a}nsch}}}, \bibinfo{journal}{Physical Review Letters}
  \textbf{\bibinfo{volume}{104}}, \bibinfo{eid}{233001} (\bibinfo{year}{2010}).

\bibitem[{\citenamefont{{Parthey} et~al.}(2011)\citenamefont{{Parthey},
  {Matveev}, {Alnis}, {Bernhardt}, {Beyer}, {Holzwarth}, {Maistrou}, {Pohl},
  {Predehl}, {Udem} et~al.}}]{2011PhRvL.107t3001P}
\bibinfo{author}{\bibfnamefont{C.~G.} \bibnamefont{{Parthey}}},
  \bibinfo{author}{\bibfnamefont{A.}~\bibnamefont{{Matveev}}},
  \bibinfo{author}{\bibfnamefont{J.}~\bibnamefont{{Alnis}}},
  \bibinfo{author}{\bibfnamefont{B.}~\bibnamefont{{Bernhardt}}},
  \bibinfo{author}{\bibfnamefont{A.}~\bibnamefont{{Beyer}}},
  \bibinfo{author}{\bibfnamefont{R.}~\bibnamefont{{Holzwarth}}},
  \bibinfo{author}{\bibfnamefont{A.}~\bibnamefont{{Maistrou}}},
  \bibinfo{author}{\bibfnamefont{R.}~\bibnamefont{{Pohl}}},
  \bibinfo{author}{\bibfnamefont{K.}~\bibnamefont{{Predehl}}},
  \bibinfo{author}{\bibfnamefont{T.}~\bibnamefont{{Udem}}},
  \bibnamefont{et~al.}, \bibinfo{journal}{Physical Review Letters}
  \textbf{\bibinfo{volume}{107}}, \bibinfo{eid}{203001} (\bibinfo{year}{2011}),
  \eprint{1107.3101}.

\bibitem[{\citenamefont{{Jentschura}}(2011{\natexlab{a}})}]{2011AnPhy.326..516%
J}
\bibinfo{author}{\bibfnamefont{U.~D.} \bibnamefont{{Jentschura}}},
  \bibinfo{journal}{Annals of Physics} \textbf{\bibinfo{volume}{326}},
  \bibinfo{pages}{516} (\bibinfo{year}{2011}{\natexlab{a}}),
  \eprint{1011.5453}.

\bibitem[{\citenamefont{{Karr} and {Hilico}}(2012)}]{2012PhRvL.109j3401K}
\bibinfo{author}{\bibfnamefont{J.-P.} \bibnamefont{{Karr}}} \bibnamefont{and}
  \bibinfo{author}{\bibfnamefont{L.}~\bibnamefont{{Hilico}}},
  \bibinfo{journal}{Physical Review Letters} \textbf{\bibinfo{volume}{109}},
  \bibinfo{eid}{103401} (\bibinfo{year}{2012}), \eprint{1205.0633}.

\bibitem[{\citenamefont{{Pohl} et~al.}(2006)\citenamefont{{Pohl}, {Daniel},
  {Hartmann}, {Hauser}, {Kottmann}, {Markushin}, {M{\"u}hlbauer}, {Petitjean},
  {Schott}, {Taqqu} et~al.}}]{2006PhRvL..97s3402P}
\bibinfo{author}{\bibfnamefont{R.}~\bibnamefont{{Pohl}}},
  \bibinfo{author}{\bibfnamefont{H.}~\bibnamefont{{Daniel}}},
  \bibinfo{author}{\bibfnamefont{F.~J.} \bibnamefont{{Hartmann}}},
  \bibinfo{author}{\bibfnamefont{P.}~\bibnamefont{{Hauser}}},
  \bibinfo{author}{\bibfnamefont{F.}~\bibnamefont{{Kottmann}}},
  \bibinfo{author}{\bibfnamefont{V.~E.} \bibnamefont{{Markushin}}},
  \bibinfo{author}{\bibfnamefont{M.}~\bibnamefont{{M{\"u}hlbauer}}},
  \bibinfo{author}{\bibfnamefont{C.}~\bibnamefont{{Petitjean}}},
  \bibinfo{author}{\bibfnamefont{W.}~\bibnamefont{{Schott}}},
  \bibinfo{author}{\bibfnamefont{D.}~\bibnamefont{{Taqqu}}},
  \bibnamefont{et~al.}, \bibinfo{journal}{Physical Review Letters}
  \textbf{\bibinfo{volume}{97}}, \bibinfo{eid}{193402} (\bibinfo{year}{2006}),
  \eprint{arXiv:physics/0610100}.

\bibitem[{\citenamefont{{I.~Eides} et~al.}(2001)\citenamefont{{I.~Eides},
  {Grotch}, and {Shelyuto}}}]{2001PhR...342...63I}
\bibinfo{author}{\bibfnamefont{M.}~\bibnamefont{{I.~Eides}}},
  \bibinfo{author}{\bibfnamefont{H.}~\bibnamefont{{Grotch}}}, \bibnamefont{and}
  \bibinfo{author}{\bibfnamefont{V.~A.} \bibnamefont{{Shelyuto}}},
  \bibinfo{journal}{Physics Report} \textbf{\bibinfo{volume}{342}},
  \bibinfo{pages}{63} (\bibinfo{year}{2001}), \eprint{arXiv:hep-ph/0002158}.

\bibitem[{\citenamefont{{Jentschura} and {Mohr}}(2004)}]{2004PhRvA..69f4103J}
\bibinfo{author}{\bibfnamefont{U.~D.} \bibnamefont{{Jentschura}}}
  \bibnamefont{and} \bibinfo{author}{\bibfnamefont{P.~J.}
  \bibnamefont{{Mohr}}}, \bibinfo{journal}{\pra} \textbf{\bibinfo{volume}{69}},
  \bibinfo{eid}{064103} (\bibinfo{year}{2004}), \eprint{arXiv:hep-ph/0405137}.

\bibitem[{\citenamefont{{Jentschura} et~al.}(2005)\citenamefont{{Jentschura},
  {Czarnecki}, and {Pachucki}}}]{2005PhRvA..72f2102J}
\bibinfo{author}{\bibfnamefont{U.~D.} \bibnamefont{{Jentschura}}},
  \bibinfo{author}{\bibfnamefont{A.}~\bibnamefont{{Czarnecki}}},
  \bibnamefont{and}
  \bibinfo{author}{\bibfnamefont{K.}~\bibnamefont{{Pachucki}}},
  \bibinfo{journal}{\pra} \textbf{\bibinfo{volume}{72}}, \bibinfo{eid}{062102}
  (\bibinfo{year}{2005}), \eprint{arXiv:physics/0603123}.

\bibitem[{\citenamefont{{Pachucki} and
  {Jentschura}}(2003)}]{2003PhRvL..91k3005P}
\bibinfo{author}{\bibfnamefont{K.}~\bibnamefont{{Pachucki}}} \bibnamefont{and}
  \bibinfo{author}{\bibfnamefont{U.~D.} \bibnamefont{{Jentschura}}},
  \bibinfo{journal}{Physical Review Letters} \textbf{\bibinfo{volume}{91}},
  \bibinfo{eid}{113005} (\bibinfo{year}{2003}), \eprint{arXiv:hep-ph/0310060}.

\bibitem[{\citenamefont{{Eides} and {Grotch}}(1997)}]{1997PhRvA..55.3351E}
\bibinfo{author}{\bibfnamefont{M.~I.} \bibnamefont{{Eides}}} \bibnamefont{and}
  \bibinfo{author}{\bibfnamefont{H.}~\bibnamefont{{Grotch}}},
  \bibinfo{journal}{\pra} \textbf{\bibinfo{volume}{55}}, \bibinfo{pages}{3351}
  (\bibinfo{year}{1997}), \eprint{arXiv:hep-ph/9611399}.

\bibitem[{\citenamefont{{Yerokhin}}(2010)}]{2010EPJD...58...57Y}
\bibinfo{author}{\bibfnamefont{V.~A.} \bibnamefont{{Yerokhin}}},
  \bibinfo{journal}{European Physical Journal D} \textbf{\bibinfo{volume}{58}},
  \bibinfo{pages}{57} (\bibinfo{year}{2010}), \eprint{1001.5436}.

\bibitem[{\citenamefont{{Borie}}(2012)}]{2012AnPhy.327..733B}
\bibinfo{author}{\bibfnamefont{E.}~\bibnamefont{{Borie}}},
  \bibinfo{journal}{Annals of Physics} \textbf{\bibinfo{volume}{327}},
  \bibinfo{pages}{733} (\bibinfo{year}{2012}), \eprint{1103.1772}.

\bibitem[{\citenamefont{{Borie}}(2005)}]{2005PhRvA..71c2508B}
\bibinfo{author}{\bibfnamefont{E.}~\bibnamefont{{Borie}}},
  \bibinfo{journal}{\pra} \textbf{\bibinfo{volume}{71}}, \bibinfo{eid}{032508}
  (\bibinfo{year}{2005}), \eprint{arXiv:physics/0410051}.

\bibitem[{\citenamefont{{Borie} and {Rinker}}(1982)}]{1982RvMP...54...67B}
\bibinfo{author}{\bibfnamefont{E.}~\bibnamefont{{Borie}}} \bibnamefont{and}
  \bibinfo{author}{\bibfnamefont{G.~A.} \bibnamefont{{Rinker}}},
  \bibinfo{journal}{Reviews of Modern Physics} \textbf{\bibinfo{volume}{54}},
  \bibinfo{pages}{67} (\bibinfo{year}{1982}).

\bibitem[{\citenamefont{{Antognini} et~al.}(2012)\citenamefont{{Antognini},
  {Kottmann}, {Biraben}, {Indelicato}, {Nez}, and
  {Pohl}}}]{2012arXiv1208.2637A}
\bibinfo{author}{\bibfnamefont{A.}~\bibnamefont{{Antognini}}},
  \bibinfo{author}{\bibfnamefont{F.}~\bibnamefont{{Kottmann}}},
  \bibinfo{author}{\bibfnamefont{F.}~\bibnamefont{{Biraben}}},
  \bibinfo{author}{\bibfnamefont{P.}~\bibnamefont{{Indelicato}}},
  \bibinfo{author}{\bibfnamefont{F.}~\bibnamefont{{Nez}}}, \bibnamefont{and}
  \bibinfo{author}{\bibfnamefont{R.}~\bibnamefont{{Pohl}}},
  \bibinfo{journal}{ArXiv e-prints}  (\bibinfo{year}{2012}),
  \eprint{1208.2637}.

\bibitem[{\citenamefont{{Jentschura}}(2011{\natexlab{b}})}]{2011AnPhy.326..500%
J}
\bibinfo{author}{\bibfnamefont{U.~D.} \bibnamefont{{Jentschura}}},
  \bibinfo{journal}{Annals of Physics} \textbf{\bibinfo{volume}{326}},
  \bibinfo{pages}{500} (\bibinfo{year}{2011}{\natexlab{b}}),
  \eprint{1011.5275}.

\bibitem[{\citenamefont{{Indelicato}}(2012)}]{2012arXiv1210.5828I}
\bibinfo{author}{\bibfnamefont{P.}~\bibnamefont{{Indelicato}}},
  \bibinfo{journal}{ArXiv e-prints}  (\bibinfo{year}{2012}),
  \eprint{1210.5828}.

\bibitem[{\citenamefont{{Carroll} et~al.}(2011)\citenamefont{{Carroll},
  {Thomas}, {Rafelski}, and {Miller}}}]{2011PhRvA..84a2506C}
\bibinfo{author}{\bibfnamefont{J.~D.} \bibnamefont{{Carroll}}},
  \bibinfo{author}{\bibfnamefont{A.~W.} \bibnamefont{{Thomas}}},
  \bibinfo{author}{\bibfnamefont{J.}~\bibnamefont{{Rafelski}}},
  \bibnamefont{and} \bibinfo{author}{\bibfnamefont{G.~A.}
  \bibnamefont{{Miller}}}, \bibinfo{journal}{\pra}
  \textbf{\bibinfo{volume}{84}}, \bibinfo{eid}{012506} (\bibinfo{year}{2011}),
  \eprint{1104.2971}.

\bibitem[{\citenamefont{{Pachucki}}(1996)}]{1996PhRvA..53.2092P}
\bibinfo{author}{\bibfnamefont{K.}~\bibnamefont{{Pachucki}}},
  \bibinfo{journal}{\pra} \textbf{\bibinfo{volume}{53}}, \bibinfo{pages}{2092}
  (\bibinfo{year}{1996}).

\bibitem[{\citenamefont{{Pachucki}}(1999)}]{1999PhRvA..60.3593P}
\bibinfo{author}{\bibfnamefont{K.}~\bibnamefont{{Pachucki}}},
  \bibinfo{journal}{\pra} \textbf{\bibinfo{volume}{60}}, \bibinfo{pages}{3593}
  (\bibinfo{year}{1999}), \eprint{arXiv:physics/9906002}.

\bibitem[{\citenamefont{{Martynenko}}(2006)}]{2006PAN....69.1309M}
\bibinfo{author}{\bibfnamefont{A.~P.} \bibnamefont{{Martynenko}}},
  \bibinfo{journal}{Physics of Atomic Nuclei} \textbf{\bibinfo{volume}{69}},
  \bibinfo{pages}{1309} (\bibinfo{year}{2006}), \eprint{arXiv:hep-ph/0509236}.

\bibitem[{\citenamefont{{Hill} and {Paz}}(2011)}]{2011PhRvL.107p0402H}
\bibinfo{author}{\bibfnamefont{R.~J.} \bibnamefont{{Hill}}} \bibnamefont{and}
  \bibinfo{author}{\bibfnamefont{G.}~\bibnamefont{{Paz}}},
  \bibinfo{journal}{Physical Review Letters} \textbf{\bibinfo{volume}{107}},
  \bibinfo{eid}{160402} (\bibinfo{year}{2011}), \eprint{1103.4617}.

\bibitem[{\citenamefont{{Miller} et~al.}(2011)\citenamefont{{Miller}, {Thomas},
  {Carroll}, and {Rafelski}}}]{2011PhRvA..84b0101M}
\bibinfo{author}{\bibfnamefont{G.~A.} \bibnamefont{{Miller}}},
  \bibinfo{author}{\bibfnamefont{A.~W.} \bibnamefont{{Thomas}}},
  \bibinfo{author}{\bibfnamefont{J.~D.} \bibnamefont{{Carroll}}},
  \bibnamefont{and}
  \bibinfo{author}{\bibfnamefont{J.}~\bibnamefont{{Rafelski}}},
  \bibinfo{journal}{\pra} \textbf{\bibinfo{volume}{84}}, \bibinfo{eid}{020101}
  (\bibinfo{year}{2011}), \eprint{1101.4073}.

\bibitem[{\citenamefont{{Miller} et~al.}(2012)\citenamefont{{Miller}, {Thomas},
  and {Carroll}}}]{2012PhRvC..86f5201M}
\bibinfo{author}{\bibfnamefont{G.~A.} \bibnamefont{{Miller}}},
  \bibinfo{author}{\bibfnamefont{A.~W.} \bibnamefont{{Thomas}}},
  \bibnamefont{and} \bibinfo{author}{\bibfnamefont{J.~D.}
  \bibnamefont{{Carroll}}}, \bibinfo{journal}{\prc}
  \textbf{\bibinfo{volume}{86}}, \bibinfo{eid}{065201} (\bibinfo{year}{2012}),
  \eprint{1207.0549}.

\bibitem[{\citenamefont{{Miller}}(2013)}]{2013PhLB..718.1078M}
\bibinfo{author}{\bibfnamefont{G.~A.} \bibnamefont{{Miller}}},
  \bibinfo{journal}{Physics Letters B} \textbf{\bibinfo{volume}{718}},
  \bibinfo{pages}{1078} (\bibinfo{year}{2013}), \eprint{1209.4667}.

\bibitem[{\citenamefont{{Barger} et~al.}(2011)\citenamefont{{Barger}, {Chiang},
  {Keung}, and {Marfatia}}}]{2011PhRvL.106o3001B}
\bibinfo{author}{\bibfnamefont{V.}~\bibnamefont{{Barger}}},
  \bibinfo{author}{\bibfnamefont{C.-W.} \bibnamefont{{Chiang}}},
  \bibinfo{author}{\bibfnamefont{W.-Y.} \bibnamefont{{Keung}}},
  \bibnamefont{and}
  \bibinfo{author}{\bibfnamefont{D.}~\bibnamefont{{Marfatia}}},
  \bibinfo{journal}{Physical Review Letters} \textbf{\bibinfo{volume}{106}},
  \bibinfo{eid}{153001} (\bibinfo{year}{2011}), \eprint{1011.3519}.

\bibitem[{\citenamefont{{Barger} et~al.}(2012)\citenamefont{{Barger}, {Chiang},
  {Keung}, and {Marfatia}}}]{2012PhRvL.108h1802B}
\bibinfo{author}{\bibfnamefont{V.}~\bibnamefont{{Barger}}},
  \bibinfo{author}{\bibfnamefont{C.-W.} \bibnamefont{{Chiang}}},
  \bibinfo{author}{\bibfnamefont{W.-Y.} \bibnamefont{{Keung}}},
  \bibnamefont{and}
  \bibinfo{author}{\bibfnamefont{D.}~\bibnamefont{{Marfatia}}},
  \bibinfo{journal}{Physical Review Letters} \textbf{\bibinfo{volume}{108}},
  \bibinfo{eid}{081802} (\bibinfo{year}{2012}), \eprint{1109.6652}.

\bibitem[{\citenamefont{{Brax} and {Burrage}}(2011)}]{2011PhRvD..83c5020B}
\bibinfo{author}{\bibfnamefont{P.}~\bibnamefont{{Brax}}} \bibnamefont{and}
  \bibinfo{author}{\bibfnamefont{C.}~\bibnamefont{{Burrage}}},
  \bibinfo{journal}{\prd} \textbf{\bibinfo{volume}{83}}, \bibinfo{eid}{035020}
  (\bibinfo{year}{2011}), \eprint{1010.5108}.

\bibitem[{\citenamefont{{Batell} et~al.}(2011)\citenamefont{{Batell}, {McKeen},
  and {Pospelov}}}]{2011PhRvL.107a1803B}
\bibinfo{author}{\bibfnamefont{B.}~\bibnamefont{{Batell}}},
  \bibinfo{author}{\bibfnamefont{D.}~\bibnamefont{{McKeen}}}, \bibnamefont{and}
  \bibinfo{author}{\bibfnamefont{M.}~\bibnamefont{{Pospelov}}},
  \bibinfo{journal}{Physical Review Letters} \textbf{\bibinfo{volume}{107}},
  \bibinfo{eid}{011803} (\bibinfo{year}{2011}), \eprint{1103.0721}.

\bibitem[{\citenamefont{{Carlson} and {Rislow}}(2012)}]{2012PhRvD..86c5013C}
\bibinfo{author}{\bibfnamefont{C.~E.} \bibnamefont{{Carlson}}}
  \bibnamefont{and} \bibinfo{author}{\bibfnamefont{B.~C.}
  \bibnamefont{{Rislow}}}, \bibinfo{journal}{\prd}
  \textbf{\bibinfo{volume}{86}}, \bibinfo{eid}{035013} (\bibinfo{year}{2012}),
  \eprint{1206.3587}.

\bibitem[{\citenamefont{{Tucker-Smith} and
  {Yavin}}(2011)}]{2011PhRvD..83j1702T}
\bibinfo{author}{\bibfnamefont{D.}~\bibnamefont{{Tucker-Smith}}}
  \bibnamefont{and} \bibinfo{author}{\bibfnamefont{I.}~\bibnamefont{{Yavin}}},
  \bibinfo{journal}{\prd} \textbf{\bibinfo{volume}{83}}, \bibinfo{eid}{101702}
  (\bibinfo{year}{2011}), \eprint{1011.4922}.

\bibitem[{\citenamefont{{Arkani-Hamed}
  et~al.}(1998)\citenamefont{{Arkani-Hamed}, {Dimopoulos}, and
  {Dvali}}}]{1998PhLB..429..263A}
\bibinfo{author}{\bibfnamefont{N.}~\bibnamefont{{Arkani-Hamed}}},
  \bibinfo{author}{\bibfnamefont{S.}~\bibnamefont{{Dimopoulos}}},
  \bibnamefont{and} \bibinfo{author}{\bibfnamefont{G.}~\bibnamefont{{Dvali}}},
  \bibinfo{journal}{Physics Letters B} \textbf{\bibinfo{volume}{429}},
  \bibinfo{pages}{263} (\bibinfo{year}{1998}), \eprint{arXiv:hep-ph/9803315}.

\bibitem[{\citenamefont{{Hoyle} et~al.}(2004)\citenamefont{{Hoyle}, {Kapner},
  {Heckel}, {Adelberger}, {Gundlach}, {Schmidt}, and
  {Swanson}}}]{2004PhRvD..70d2004H}
\bibinfo{author}{\bibfnamefont{C.~D.} \bibnamefont{{Hoyle}}},
  \bibinfo{author}{\bibfnamefont{D.~J.} \bibnamefont{{Kapner}}},
  \bibinfo{author}{\bibfnamefont{B.~R.} \bibnamefont{{Heckel}}},
  \bibinfo{author}{\bibfnamefont{E.~G.} \bibnamefont{{Adelberger}}},
  \bibinfo{author}{\bibfnamefont{J.~H.} \bibnamefont{{Gundlach}}},
  \bibinfo{author}{\bibfnamefont{U.}~\bibnamefont{{Schmidt}}},
  \bibnamefont{and} \bibinfo{author}{\bibfnamefont{H.~E.}
  \bibnamefont{{Swanson}}}, \bibinfo{journal}{\prd}
  \textbf{\bibinfo{volume}{70}}, \bibinfo{eid}{042004} (\bibinfo{year}{2004}),
  \eprint{arXiv:hep-ph/0405262}.

\bibitem[{\citenamefont{{Kapner} et~al.}(2007)\citenamefont{{Kapner}, {Cook},
  {Adelberger}, {Gundlach}, {Heckel}, {Hoyle}, and
  {Swanson}}}]{2007PhRvL..98b1101K}
\bibinfo{author}{\bibfnamefont{D.~J.} \bibnamefont{{Kapner}}},
  \bibinfo{author}{\bibfnamefont{T.~S.} \bibnamefont{{Cook}}},
  \bibinfo{author}{\bibfnamefont{E.~G.} \bibnamefont{{Adelberger}}},
  \bibinfo{author}{\bibfnamefont{J.~H.} \bibnamefont{{Gundlach}}},
  \bibinfo{author}{\bibfnamefont{B.~R.} \bibnamefont{{Heckel}}},
  \bibinfo{author}{\bibfnamefont{C.~D.} \bibnamefont{{Hoyle}}},
  \bibnamefont{and} \bibinfo{author}{\bibfnamefont{H.~E.}
  \bibnamefont{{Swanson}}}, \bibinfo{journal}{Physical Review Letters}
  \textbf{\bibinfo{volume}{98}}, \bibinfo{eid}{021101} (\bibinfo{year}{2007}),
  \eprint{arXiv:hep-ph/0611184}.

\bibitem[{\citenamefont{{Sushkov} et~al.}(2011)\citenamefont{{Sushkov}, {Kim},
  {Dalvit}, and {Lamoreaux}}}]{2011PhRvL.107q1101S}
\bibinfo{author}{\bibfnamefont{A.~O.} \bibnamefont{{Sushkov}}},
  \bibinfo{author}{\bibfnamefont{W.~J.} \bibnamefont{{Kim}}},
  \bibinfo{author}{\bibfnamefont{D.~A.~R.} \bibnamefont{{Dalvit}}},
  \bibnamefont{and} \bibinfo{author}{\bibfnamefont{S.~K.}
  \bibnamefont{{Lamoreaux}}}, \bibinfo{journal}{Physical Review Letters}
  \textbf{\bibinfo{volume}{107}}, \bibinfo{eid}{171101} (\bibinfo{year}{2011}).

\bibitem[{\citenamefont{{Yang} et~al.}(2012)\citenamefont{{Yang}, {Zhan},
  {Wang}, {Shao}, {Tu}, {Tan}, and {Luo}}}]{2012PhRvL.108h1101Y}
\bibinfo{author}{\bibfnamefont{S.-Q.} \bibnamefont{{Yang}}},
  \bibinfo{author}{\bibfnamefont{B.-F.} \bibnamefont{{Zhan}}},
  \bibinfo{author}{\bibfnamefont{Q.-L.} \bibnamefont{{Wang}}},
  \bibinfo{author}{\bibfnamefont{C.-G.} \bibnamefont{{Shao}}},
  \bibinfo{author}{\bibfnamefont{L.-C.} \bibnamefont{{Tu}}},
  \bibinfo{author}{\bibfnamefont{W.-H.} \bibnamefont{{Tan}}}, \bibnamefont{and}
  \bibinfo{author}{\bibfnamefont{J.}~\bibnamefont{{Luo}}},
  \bibinfo{journal}{Physical Review Letters} \textbf{\bibinfo{volume}{108}},
  \bibinfo{eid}{081101} (\bibinfo{year}{2012}).

\bibitem[{\citenamefont{Geraci et~al.}(2008)\citenamefont{Geraci, Smullin,
  Weld, Chiaverini, and Kapitulnik}}]{PhysRevD.78.022002}
\bibinfo{author}{\bibfnamefont{A.~A.} \bibnamefont{Geraci}},
  \bibinfo{author}{\bibfnamefont{S.~J.} \bibnamefont{Smullin}},
  \bibinfo{author}{\bibfnamefont{D.~M.} \bibnamefont{Weld}},
  \bibinfo{author}{\bibfnamefont{J.}~\bibnamefont{Chiaverini}},
  \bibnamefont{and}
  \bibinfo{author}{\bibfnamefont{A.}~\bibnamefont{Kapitulnik}},
  \bibinfo{journal}{Phys. Rev. D} \textbf{\bibinfo{volume}{78}},
  \bibinfo{pages}{022002} (\bibinfo{year}{2008}).

\bibitem[{\citenamefont{{Klimchitskaya}
  et~al.}(2012)\citenamefont{{Klimchitskaya}, {Mohideen}, and
  {Mostepanenko}}}]{2012PhRvD..86f5025K}
\bibinfo{author}{\bibfnamefont{G.~L.} \bibnamefont{{Klimchitskaya}}},
  \bibinfo{author}{\bibfnamefont{U.}~\bibnamefont{{Mohideen}}},
  \bibnamefont{and} \bibinfo{author}{\bibfnamefont{V.~M.}
  \bibnamefont{{Mostepanenko}}}, \bibinfo{journal}{\prd}
  \textbf{\bibinfo{volume}{86}}, \bibinfo{eid}{065025} (\bibinfo{year}{2012}),
  \eprint{1209.0086}.

\bibitem[{\citenamefont{{Li} et~al.}(2008)\citenamefont{{Li}, {Ni}, and
  {Antonio Pulido}}}]{2008ChPhB..17...70L}
\bibinfo{author}{\bibfnamefont{Z.-G.} \bibnamefont{{Li}}},
  \bibinfo{author}{\bibfnamefont{W.-T.} \bibnamefont{{Ni}}}, \bibnamefont{and}
  \bibinfo{author}{\bibfnamefont{P.}~\bibnamefont{{Antonio Pulido}}},
  \bibinfo{journal}{Chinese Physics B} \textbf{\bibinfo{volume}{17}},
  \bibinfo{pages}{70} (\bibinfo{year}{2008}), \eprint{0706.3961}.

\bibitem[{\citenamefont{{Adelberger} et~al.}(2007)\citenamefont{{Adelberger},
  {Heckel}, {Hoedl}, {Hoyle}, {Kapner}, and {Upadhye}}}]{2007PhRvL..98m1104A}
\bibinfo{author}{\bibfnamefont{E.~G.} \bibnamefont{{Adelberger}}},
  \bibinfo{author}{\bibfnamefont{B.~R.} \bibnamefont{{Heckel}}},
  \bibinfo{author}{\bibfnamefont{S.}~\bibnamefont{{Hoedl}}},
  \bibinfo{author}{\bibfnamefont{C.~D.} \bibnamefont{{Hoyle}}},
  \bibinfo{author}{\bibfnamefont{D.~J.} \bibnamefont{{Kapner}}},
  \bibnamefont{and}
  \bibinfo{author}{\bibfnamefont{A.}~\bibnamefont{{Upadhye}}},
  \bibinfo{journal}{Physical Review Letters} \textbf{\bibinfo{volume}{98}},
  \bibinfo{eid}{131104} (\bibinfo{year}{2007}), \eprint{arXiv:hep-ph/0611223}.

\bibitem[{\citenamefont{{Lambrecht} et~al.}(2011)\citenamefont{{Lambrecht},
  {Canaguier-Durand}, {Gu{\'e}rout}, and {Reynaud}}}]{2011LNP...834...97L}
\bibinfo{author}{\bibfnamefont{A.}~\bibnamefont{{Lambrecht}}},
  \bibinfo{author}{\bibfnamefont{A.}~\bibnamefont{{Canaguier-Durand}}},
  \bibinfo{author}{\bibfnamefont{R.}~\bibnamefont{{Gu{\'e}rout}}},
  \bibnamefont{and}
  \bibinfo{author}{\bibfnamefont{S.}~\bibnamefont{{Reynaud}}}, in
  \emph{\bibinfo{booktitle}{Lecture Notes in Physics, Berlin Springer Verlag}},
  edited by \bibinfo{editor}{\bibfnamefont{D.}~\bibnamefont{{Dalvit}}},
  \bibinfo{editor}{\bibfnamefont{P.}~\bibnamefont{{Milonni}}},
  \bibinfo{editor}{\bibfnamefont{D.}~\bibnamefont{{Roberts}}},
  \bibnamefont{and} \bibinfo{editor}{\bibfnamefont{F.}~\bibnamefont{{da Rosa}}}
  (\bibinfo{year}{2011}), vol. \bibinfo{volume}{834} of
  \emph{\bibinfo{series}{Lecture Notes in Physics, Berlin Springer Verlag}},
  p.~\bibinfo{pages}{97}, \eprint{1006.2959}.

\bibitem[{\citenamefont{{Bird}}(2008)}]{2008PhDT.......280B}
\bibinfo{author}{\bibfnamefont{C.~S.} \bibnamefont{{Bird}}},
  \bibinfo{journal}{PhD. Thesis, University of Victoria, Canada}
  (\bibinfo{year}{2008}).

\bibitem[{\citenamefont{{Machado} et~al.}(2011)\citenamefont{{Machado},
  {Nunokawa}, and {Funchal}}}]{2011PhRvD..84a3003M}
\bibinfo{author}{\bibfnamefont{P.~A.~N.} \bibnamefont{{Machado}}},
  \bibinfo{author}{\bibfnamefont{H.}~\bibnamefont{{Nunokawa}}},
  \bibnamefont{and} \bibinfo{author}{\bibfnamefont{R.~Z.}
  \bibnamefont{{Funchal}}}, \bibinfo{journal}{\prd}
  \textbf{\bibinfo{volume}{84}}, \bibinfo{eid}{013003} (\bibinfo{year}{2011}),
  \eprint{1101.0003}.

\bibitem[{\citenamefont{{Fermi-LAT Collaboration}
  et~al.}(2012)\citenamefont{{Fermi-LAT Collaboration}, {Ajello}, {Baldini},
  {Barbiellini}, {Bastieri}, {Bechtol}, {Bellazzini}, {Berenji}, {Bloom},
  {Bonamente} et~al.}}]{2012JCAP...02..012F}
\bibinfo{author}{\bibnamefont{{Fermi-LAT Collaboration}}},
  \bibinfo{author}{\bibfnamefont{M.}~\bibnamefont{{Ajello}}},
  \bibinfo{author}{\bibfnamefont{L.}~\bibnamefont{{Baldini}}},
  \bibinfo{author}{\bibfnamefont{G.}~\bibnamefont{{Barbiellini}}},
  \bibinfo{author}{\bibfnamefont{D.}~\bibnamefont{{Bastieri}}},
  \bibinfo{author}{\bibfnamefont{K.}~\bibnamefont{{Bechtol}}},
  \bibinfo{author}{\bibfnamefont{R.}~\bibnamefont{{Bellazzini}}},
  \bibinfo{author}{\bibfnamefont{B.}~\bibnamefont{{Berenji}}},
  \bibinfo{author}{\bibfnamefont{E.~D.} \bibnamefont{{Bloom}}},
  \bibinfo{author}{\bibfnamefont{E.}~\bibnamefont{{Bonamente}}},
  \bibnamefont{et~al.}, \bibinfo{journal}{\jcap} \textbf{\bibinfo{volume}{2}},
  \bibinfo{eid}{012} (\bibinfo{year}{2012}).

\bibitem[{\citenamefont{{Aad} et~al.}(2013)\citenamefont{{Aad}, {Abajyan},
  {Abbott}, {Abdallah}, {Abdel Khalek}, {Abdelalim}, {Abdinov}, {Aben}, {Abi},
  {Abolins} et~al.}}]{2013PhRvL.110a1802A}
\bibinfo{author}{\bibfnamefont{G.}~\bibnamefont{{Aad}}},
  \bibinfo{author}{\bibfnamefont{T.}~\bibnamefont{{Abajyan}}},
  \bibinfo{author}{\bibfnamefont{B.}~\bibnamefont{{Abbott}}},
  \bibinfo{author}{\bibfnamefont{J.}~\bibnamefont{{Abdallah}}},
  \bibinfo{author}{\bibfnamefont{S.}~\bibnamefont{{Abdel Khalek}}},
  \bibinfo{author}{\bibfnamefont{A.~A.} \bibnamefont{{Abdelalim}}},
  \bibinfo{author}{\bibfnamefont{O.}~\bibnamefont{{Abdinov}}},
  \bibinfo{author}{\bibfnamefont{R.}~\bibnamefont{{Aben}}},
  \bibinfo{author}{\bibfnamefont{B.}~\bibnamefont{{Abi}}},
  \bibinfo{author}{\bibfnamefont{M.}~\bibnamefont{{Abolins}}},
  \bibnamefont{et~al.}, \bibinfo{journal}{Physical Review Letters}
  \textbf{\bibinfo{volume}{110}}, \bibinfo{eid}{011802} (\bibinfo{year}{2013}).

\bibitem[{\citenamefont{{ATLAS Collaboration}}(2012)}]{2012arXiv1211.1150A}
\bibinfo{author}{\bibnamefont{{ATLAS Collaboration}}}, \bibinfo{journal}{ArXiv
  e-prints}  (\bibinfo{year}{2012}), \eprint{1211.1150}.

\end{thebibliography}

\end{document}